\documentclass[11pt]{elsart}
\usepackage{epsfig}
\usepackage{amssymb}

\def\l{\ell}
\newcommand{\cl}{C_{\l}}
\def\be{\begin{equation}}
\def\ee{\end{equation}}

\begin{document}

\begin{frontmatter}

\title{{\bf Galactic Magnetic Turbulence from Radio data}}
\thanks{Preprint number: DFTT 3/2011}

\author{Marco Regis}
\ead{regis@to.infn.it}

\address{{\small {\it Dipartimento di Fisica Teorica, Universit\`a di Torino} \\
{\it Istituto Nazionale di Fisica Nucleare, Sezione di Torino} \\
{\it via P. Giuria 1, I--10125 Torino, Italy}}}

\begin{abstract}
Fluctuations in the Galactic synchrotron emission can be traced by the angular power spectrum of radio maps at low multipoles.
At frequencies below few GHz, large-scale anisotropies are mainly induced by magnetic field turbulence, since non-thermal electrons 
radiating at these frequencies are uniformly distributed over the scales of magnetic field inhomogeneities. 
By performing an analysis of five radio maps, we extract constraints on turbulence spectral index and halo scale. 
Results favour a power spectrum significantly flatter than for 3D Kolmogorov-like turbulence, and a thin halo.
This can be interpreted as an indication supporting non-conventional models of propagation of cosmic-ray particles in the Galaxy, or as a suggestion of a spectral-index break in the observed magnetic turbulence power spectrum.
\end{abstract}

\begin{keyword}
Cosmic-rays, Galactic magnetic field, Radio synchrotron emission.
\PACS{98.38.Am, 98.70.Sa, 95.85.Bh, 95.30.Qd, 98.35.Eg}
\end{keyword}

\end{frontmatter}

\section{Introduction}
One of the most important processes in the propagation of high-energy charged particles in the interstellar medium is their interaction with magnetic fields.
The scattering of cosmic-ray (CR) particles on random hydromagnetic waves leads to their effective confinement in the Galaxy.
This process of diffusion has a resonant character and it is mainly driven by the energy density associated to the random component of the magnetic field at the resonant wave number of the scattering which is given by the gyro-radius of the particle~\cite{Berezinskii:1990}.
The turbulence power spectrum at such scales and the extent of the turbulent region are key quantities in the description of CR propagation.

The aforementioned scatterings change the energy of CR, reshaping their injection energy spectra into the ones we observe at Earth.
The ratio of stable secondary to primary CR (where the reference ratio is often $B/C$, given that Boron is almost entirely secondary)
is a very useful diagnostic for such reshaping and in turn for the spectral index of turbulence power spectrum (for a recent review on CR, see, e.g.,~\cite{Strong:2007nh}), which is very often assumed to be a power-law $P(k)\sim k^{-\alpha}$ of the wavenumber $k$.
Unstable secondaries are, on the other hand, a potentially very good tracer of spatial boundaries of the turbulent halo.
Indeed, the ratio between stable and decaying isotopes depends on the CR confinement time, which is tightly related to the halo height (and the diffusion coefficient).

In this paper, we aim to derive these two key quantities (turbulence spectral index and halo height) from a different perspective.
We analyse the angular power spectrum of five radio maps. Their main component is given by the Galactic diffuse synchrotron emission.
As we will discuss in Sec.~3, Galactic electrons involved in synchrotron radiation at frequencies $\lesssim$ GHz are uniformly distributed over the scales of magnetic field inhomogeneities, given their much larger diffusive scale. 
Therefore the spatial spectrum of synchrotron intensity reflects the spectrum of magnetic fluctuations.
By restricting to maps at low radio frequencies, we are confident to trace it avoiding contamination from electrons fluctuations, and to be in a regime where synchrotron emission dominates over other diffuse contributions (e.g., dust or free-free emissions)~\cite{Gold:2010fm}.

For CR, we consider a picture of fast particles scattering in a medium with weak hydromagnetic turbulence and we will make the simplifying assumption of isotropic turbulence. The quasi-linear approximation leads to a diffusion coefficient $D\propto E^\gamma$, with $\gamma=4-\alpha$~\cite{Berezinskii:1990}, so its energy dependence is given by the same spectral index of hydromagnetic turbulence, which is derived by computing the angular spectrum of synchrotron emission and assuming a single power-law for the turbulence power-spectrum at all scales. 

The angular correlation, however, reflects statistics of turbulence only for sufficiently small angles, above which it follows a universal law $\cl\sim 1/\l$ (see, e.g.,~\cite{Chepurnov:2002vj}). 
Indeed, when the observation angle $\theta$ for an observer located 
within an homogeneous turbulence region becomes large, only points along of the 
lines-of-sight very close to the observer are correlated while farthest 
points become uncorrelated. So as the angle increases the angular correlation $w$
gets less and less contributions along line of sight, and follows the law 
$w\propto \theta^{-1}$ which leads to $C_\ell\propto \ell^{-1}$ (see, e.g., Refs.~\cite{Chepurnov:2002vj,Cho:2010kw}).
The angle of transition between the two regimes is related to the ratio between the outer scale of turbulence $L$ and the size of the turbulent region $d_{max}$. Provided an estimate for $L$ (assumed to follow from scale of energy injection from supernovae remnant), $d_{max}$ can be derived and represents an estimate for the height of the confinement halo in the Galaxy.  

Although all data considered in this paper come from single-dish surveys,
next generation of radio interferometers with wide field of view and 
a primary beam (at low frequencies) overlapping with the `large' angular scales
considered here, such as LOFAR~\cite{LOFAR}, ASKAP~\cite{ASKAP}, and SKA~\cite{SKA},
can provide promising insight into the topic. 
Previous works on angular power spectrum of radio maps and its link with Galactic synchrotron emission include~\cite{Tegmark:1995pn,Bouchet:1999gq,Tegmark:1999ke,Tucci:2000ix,Baccigalupi:2000pm,Giardino,LaPorta:2006cy,LaPorta:2008ag,Cho:2010kw,Chen:2004xx}. Their main focus has been on estimates of Galactic foreground and discussing its possible removal in CMB studies.
In this paper, we instead focus the analysis on the extraction of turbulence properties and on their links to CR models. We also extend the determination of $\cl$'s to few more maps.

In Sec.~2, we describe datasets used and summarize the computation of $\cl$. In Sec.~3, assumptions and model of angular power spectrum are discussed. Results on turbulence properties and their link to CR studies are presented in Sec.~4. Sec.~5 concludes.

\section{Data}
Numerous radio datasets concerning Galactic observations are currently available.
We followed three main criteria to choose the maps used in this work: frequency $\lesssim$ GHz, good sky coverage, and, possibly, good angular resolution. 
They include surveys of total intensity at 22, 45, 408, 820, and 1420 MHz, and of polarization at 1420 MHz. Details are summarized in Table 1 and maps are shown in Fig.1 in the HEALPix~\cite{Gorski:2004by} format.\footnote{We acknowledge MPIfR's Survey Sampler~\cite{BONN} from which some of the datasets were downloaded.} 
In order to avoid spurious projection effects, they have been obtained after regridding original maps into a much finer grid before filling pixels in the HEALPix tessellation scheme (with final linear size of pixels close to the original resolution of the survey).

The $\cl$ coefficients of the two-point angular correlation function are defined through the canonical relations $T(\theta,\phi)=\sum_{\l m}a_{\l m}Y_{\l m}(\theta,\phi)$ and $\cl=1/(2\l +1)\sum^\l_{-\l}|a_{\l m}|^2$.

The `real' power spectrum coefficients can be estimated from measured $\cl^{map}$ through $\cl=(\cl^{map}-\cl^{noise})/W $, which take into account effects of instrumental noise in $\cl^{noise}=4\,\pi\,\sigma_{noise}/N_{pix}$ (with $\sigma_{noise}$ and $N_{pix}$ being, respectively, the rms temperature noise and number of pixels of the map) and of window function $W=\exp[-\l \,(\l+1)\,\sigma^2_{beam}]$ (with $\sigma_{beam}=FWHM/\sqrt{8\,\ln{2}}$).
The statistical uncertainty in the power spectrum can be estimated as $\delta \cl=(\cl+\cl^{noise}/W)/\sqrt{(2\l+1)/2\,\Delta \l\,f_{sky}}$, where $\Delta \l$ is the number of binned multipole and $f_{sky}$ is the fraction of sky observed.

The $\cl^{map}$ for maps listed in Table~1 are shown in central and right panels of Fig.~1. They have been computed using the \verb Anafast  tool of the HEALPix package~\cite{Gorski:2004by}. 
We consider angular power spectra from all observed sky and from high-latitudes only ($b>20^{\circ}$, namely, cutting off emission from the Galactic plane).
We then binned the interval $\l=[5,100]$ in 20 bins of constant width in logarithmic scale.

\begin{table}[t]
\begin{center}
\begin{tabular}{|c|c|c|c|c|c|}
\hline
Frequency&Beamwidth&rms Noise&Fraction of Sky&$\l_{max}$&Survey\\
MHz&FWHM ($^\circ$)& K & observed& &   
\tabularnewline
\hline
\hline
22&1.1 $\times$ 1.7&5000&73\%& 55 (30) &DRAO~\cite{DRAO:22}
\tabularnewline
\hline
\hline
45&3.6&3500&86\%&17 (n.c.) &Guzman et al.~\cite{CHILE}
\tabularnewline
\hline
\hline
408&0.85&0.8&100\%&100 (100) &Haslam et al.~\cite{Haslam}
\tabularnewline
\hline
\hline
820&0.97&1.4&51\%&55 (30) &Dwingeloo~\cite{DWING}
\tabularnewline
\hline
\hline
1420  (I)&0.6&0.02&60\%&100 (100)& Stockert~\cite{Stockert} \\ 
1420  (PI)&0.6&0.015&66\%&100 (100) &DRAO~\cite{DRAO:1420}\\
1420  (PI)&0.6&0.012&44\%&100 (100) &Villa Elisa~\cite{VILLA}
\tabularnewline
\hline
\end{tabular}
\end{center}
\caption{Main parameters of surveys analysed in this work. $\l_{max}=\rm{min}(100,\l_N)$ is reported for all observed sky (and $|b|>20$), where $\l_N$ is the multipole at which noise starts dominating, see text for details. At 1420 MHz, both total (I) and polarized (PI) intensity were considered.}
\end{table}

\begin{figure}[t]
   \centering
   \includegraphics[width=\textwidth]{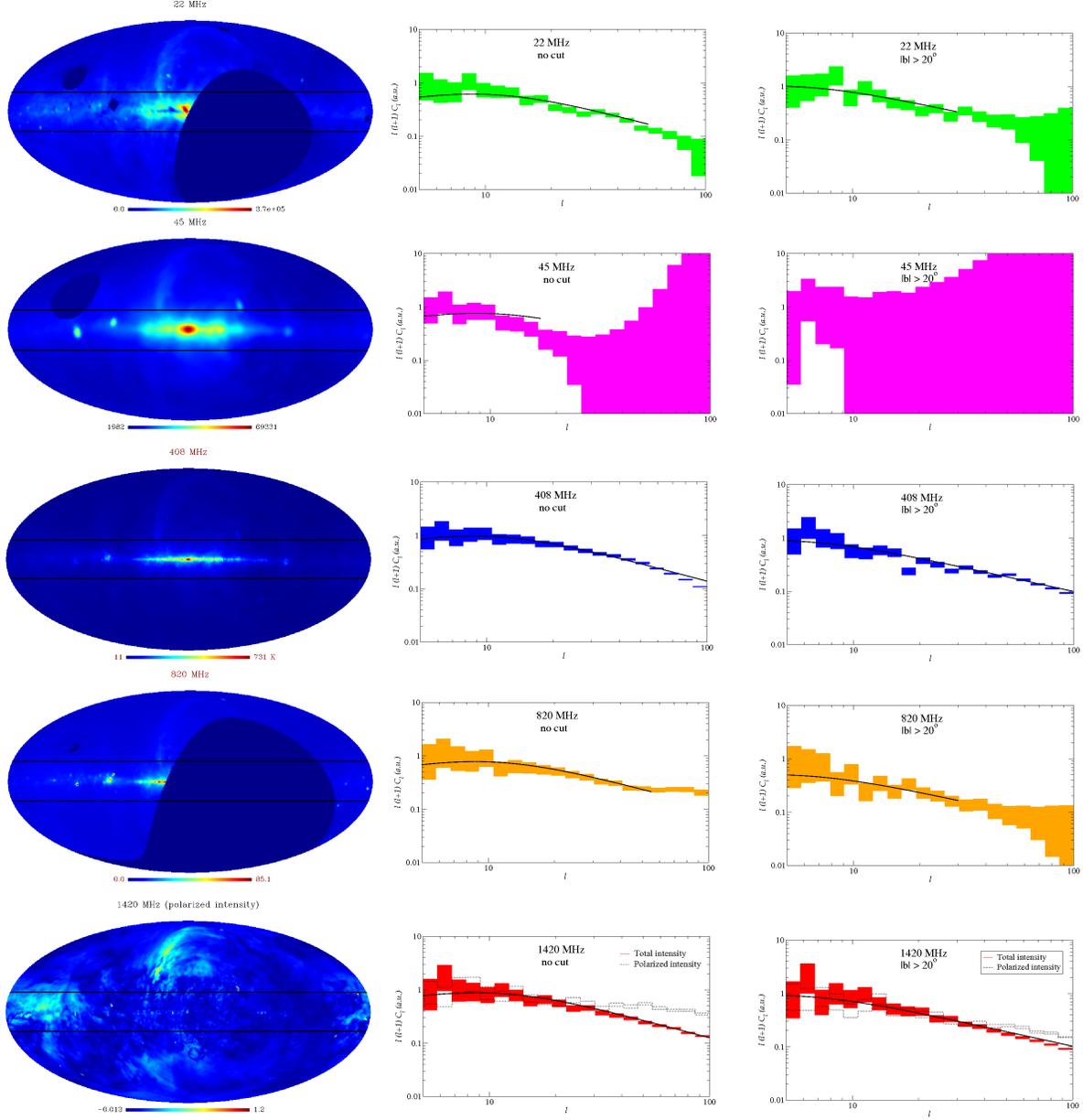}
    \caption{{\it Left Panel:} Temperature maps (in K) of Table 1 in HEALPix format. At 1420 MHz only polarized intensity (obtained combining surveys of \cite{DRAO:1420} and \cite{VILLA}) is shown. Horizontal lines indicate $|b|=20^\circ$.
{\it Middle Panel:} 1-$\sigma$ regions of (binned) angular power spectrum $\cl^{map}$ for maps in Table 1. The normalization has been arbitrarily chosen (i.e., first bin is normalized to unity) to easy comparisons. The best-fit $\cl^{turb}$ up to $\l_{max}$ is overlying (black solid curve). Note that it is obtained by fitting $(\cl^{map}-\cl^{noise})/W $ with $\cl^{turb}+\cl^s$, so it's not a direct fit to the shown observational $\cl^{map}$, although strongly related.
{\it Right Panel:} Same of middle panel, but cutting low latitudes ($|b_{cut}|=20^\circ$) data.}
\label{fig:mapscls}
 \end{figure} 

\section{Method}
In the monochromatic approximation of synchrotron radiation, the energy of an electron emitting at frequency $\nu$ is given by $E\simeq 15\sqrt{\nu_{GHz}/B_{\mu G}}$ GeV (where $\nu_{GHz}$ is the frequency in GHz and $B_{\mu G}$ is the magnetic field in $\mu G$). It follows that, assuming a Galactic magnetic field of few microGauss~\cite{Han:2009ts}, emissions at $\nu\lesssim$~GHz are mostly generated by electrons with energy significantly smaller than 10 GeV.
The mean distance diffused by such electrons before losing most of their energy can be estimated as $d_L=\sqrt{4\,D\,\tau}$~\cite{Berezinskii:1990}, where $D$ is the so-called diffusion coefficient and $\tau$ is the the time-scale for the energy loss associated to radiative processes. In the energy range of interest, typical numbers for the Galaxy are $D\sim D_0 \simeq 3\cdot 10^{28} \rm{cm}^2\rm{s}^{-1}$ and $\tau\simeq 10^{16} (E/\rm{GeV})^{-1}$ s~\cite{Strong:2007nh}, which translate into $d_L\gtrsim 4$ kpc for electrons with energy below 10 GeV.
This order of magnitude estimate tells us that the scale of isotropy of the Galactic non-thermal electron population involved in the diffuse synchrotron emission detected in the radio maps is very large. 
Note that this is also because the storage time is far larger than 
CR electron injection rate which is given by supernovae explosions 
occurring as, roughly, few per century, which means that relative 
anisotropy generated by new energy injection is small compared to 
the total population of electrons (at low energy).
For a numerical simulation of spatial distribution of Galactic CR electrons
see Sec.~4.1 in~\cite{Moskalenko:2004vh}, which agrees with above arguments. 
CR electron fluctuations are negligible and does not contribute to the angular power spectrum (except for the very first multipoles $\l\lesssim 5$).

Therefore fluctuations in the Galactic synchrotron emission can be totally ascribed to fluctuations in the magnetic field.
Throughout the paper, we will make the assumption that the latter dominate the angular power spectrum for $\l>5$ ($\theta\leq30^{\circ}$).
Fluctuations might be not all of stochastic nature, but with a contribution from the coherent magnetic field; on the other hand, in this case, fluctuations are expected to show up only at very large scales. Although synchrotron brightness does not directly trace the total magnetic field but only the component orthogonal to the line of sight, and a change in our viewing angle with respect to the smooth large-scale field (e.g., as our line of sight crosses a spiral arm) can produce a sudden brightness variation, it is unlikely that such fluctuations can contribute to the angular correlation at small scales.
The angular range discussed here can be considered to be descriptive of a truly turbulent regime.\footnote{Throughout the paper, we make the assumption that small-scale magnetic field fluctuations trace the
underlying interstellar turbulent motions, and so that measurements
of the former (static quantity) provide information on the
latter (dynamical phenomenon).}

We restrict to maps at $\nu\lesssim$ GHz, disregarding, e.g., map of \cite{Jonas} at 2.3 GHz or WMAP~\cite{WMAP} and Planck~\cite{Planck} maps, where the energy of emitting electrons would be larger and the associated mean diffusion length shorter, so fluctuations in the electron population could be sizable at smaller scales.
This effect can also introduce a frequency dependence in the $\cl$'s, since the energy and, in turn, covered distances of emitting electrons are different at different frequency. On the contrary, for the frequency range considered here, we can fit all the datasets simultaneously, exactly because anisotropies are given by the same physical framework (i.e., magnetic turbulence) at any wavelength.
We note that this implies that naive extrapolations of results found in the following Section to higher frequency (e.g., relevant for CMB studies) should be taken with a grain of salt, since they can be highly non-trivial.
 
Note also that the non-thermal electrons should not be confused with the Galactic thermal electron population which instead shows turbulences (as we will mention in the next Section), but is not relevant for synchrotron emission; it can actually introduce anisotropies through free-free absorption of radio waves~\cite{Rybicki} but this effect is important only at $\nu<10$ MHz. For the diffuse emission under investigation, synchrotron self-absorption is also not relevant. 

Both theory and numerical simulations support the idea of self-similarity of turbulence~\cite{Berezinskii:1990}, which means that the spectrum of turbulence is typically predicted to follow a power-law $k^{-\alpha}$, where $\alpha=11/3$ in the 3D Kolmogorov case.
For this kind of spatial fluctuation spectrum, the angular power spectrum can be simply modeled~\cite{Chepurnov:2002vj,Cho:2010kw} (for $\alpha > 1$). 
Indeed turbulence scaling as $k^{-\alpha}$ implies $\cl\propto \l^{-\alpha}$ in the small angle limit ($\theta<\theta_0$) and $\cl\propto \l^{-1}$ in the large angle limt ($\theta>\theta_0$), where $\theta=\pi/\l$ and the critical angle $\theta_0$ can be estimated by $\theta_0 \sim L/d_{max}$ with $L$ being the outer scale of turbulence and $d_{max}$ the boundary scale of the turbulent region.
Although this assumes homogeneous turbulence and $d_{max}$ is, strictly speaking, the distance from the center to the farthest turbulence in a spherical scenario~\cite{Cho:2010kw}, one can take it as a rough estimate of the height of the turbulent diffusive halo; this is especially true when considering $\cl$'s from high-latitudes. Moreover, since, once again, inhomogeneity in synchrotron radiation are expected to be associated to magnetic inhomogeneity only, $d_{max}$ should trace where magnetic turbulence density eventually drops, unless the turbulence has a peculiar inhomogeneity pattern.

We model the angular power spectrum of turbulence as $\cl^{turb}=\cl^{(1)}\cl^{(\alpha)}/(\cl^{(1)}+\cl^{(\alpha)})$ where $\cl^{(1)}=c_1 \l^{-1}$ and $\cl^{(\alpha)}=c_2  \l^{-\alpha}$ (namely, the two relevant scalings introduced above and in Sec.~1). Such form for $\cl^{turb}$ has been arbitrarily chosen to include scalings $\l^{-1}$ at small $\l$ and $\l^{-\alpha}$ at large $\l$. The exact shape at the transition between the two regimes affects only very mildly our results.

At frequencies below 1 GHz, synchrotron contribution largely dominates the radio diffuse emission and other components as thermal bremsstrahlung and dust can be neglected~\cite{Tegmark:1999ke}. We will however mainly refer to results obtained cutting off the Galactic plane such that this assumption can be more confidently trusted.

At high multipoles the contribution of (extra-galactic) sources becomes important and needs to be included. On the other hand, in radio maps, it is typically found to be subdominant up to $\l \lesssim 100$~\cite{Giardino,LaPorta:2008ag}, so marginally relevant for our purposes.
At $\l\sim100$ (namely, in the multipole range where they can be possibly relevant), both observations~\cite{Giardino,LaPorta:2008ag} and theoretical models~\cite{Franceschini,Zhang:2008rs} suggest the contribution of sources to be approximately flat (the `Poisson noise' term dominates).
We do not attempt to perform the delicate procedure of removing single sources, but rather we model the component, by adding an unknown flat contribution  $\cl^{s}\sim c_s=const$ at each frequency (and marginalizing over it when fitting the $\cl$'s).

Polarized intensity is potentially an even better tracer of synchrotron anisotropies than total intensity since it is not contaminated by thermal emissions. 
However, Faraday rotation effects around the source and along the line-of-sight (together with instrumental effects related to beamwidth and bandwidth of observations) can spoil the original power spectrum in a hardly predictable way. 
Faraday rotation is extremely important for radio data (while nearly irrelevant at higher frequencies) and multi-wavelength observations would be in order to single it out, but unfortunately they are not available.
From Fig.1 (bottom panel), one can see that $\cl$'s from polarized intensity shows a slightly flatter trend than total intensity.
If this is interpreted as purely due to Faraday depolarization effects, an estimate for scaling of the mean differential rotation of polarization angle $\Delta \phi$ can be derived~\cite{LaPorta:2006cy}. We obtain $(\sin{\Delta \phi}/\phi)^2\propto \cl^{PI}/\cl^I\propto \l^\mu$ with $\mu\sim 0.27$ and $0.19$ for all observed sky and high latitudes, respectively. Such a depolarization scaling (increasing with the angular scale) suggests an effect analogous to a beam depolarization (i.e., due to averaging of polarization vectors of different orientation); it is however difficult to separate different depolarization effects on $\cl$'s, including that of, for example, fluctuations in the thermal electron population.
Due to such uncertainties in extracting the intrinsic angular spectrum, we disregard polarized datasets in the fit of turbulence properties.

Theoretical $\cl$ are thus given by $\cl^{turb}+\cl^s$ which are described by four parameters $(\alpha,c_1,c_2,c_s)$ as mentioned above; they have been determined by fitting $\cl$ to the expected $(\cl^{map}-\cl^{noise})/W$ binned in the same way as $\cl^{map}$ in Fig.~1. In particular, we are interested in two parameters: $\alpha$ and $\l_0$, which, in our model, do not depend on frequency. The transition scale $\l_0 \sim \pi\,d_{max}/L$ between the two regimes $\l^{-\alpha}$ and $\l^{-1}$ is defined by $C_{\l_0}^{(1)}=C_{\l_0}^{(\alpha)}$, and can be easily expressed in terms of $c_1$ and $c_2$. The remaining two absolute normalizations of $\cl^{turb}$ and $\cl^s$ can be seen as nuisance parameters (and can vary with frequency).

For each map, we consider multipoles up to $\l_{max}=\rm{min}(100,\l_N)$ (reported in Table 1) where $\l_N$ is the multipole such that $\cl^{map} \sim \cl^{noise}/W$. 
In this way, we conservatively avoid to aggressively extract information from multipoles where spurious effects and artifacts (related to projection, resolution, point sources, etc.) might have an impact on the results.\footnote{The Haslam et al. map~\cite{Haslam} is known to show scanning artifacts along lines of constant right ascension. We downloaded the map from NASA's LAMBDA website where original data were processed to mitigate baseline striping. In any case, $\cl$ extracted from the original map or from a destriped version~\cite{Platania:2003ta} shows negligible differences~\cite{LaPorta:2008ag}.}
Indeed `re-projection' effect can affect only high multipoles, namely scales close to the angular resolution of the survey (unless data have been irreversibly binned to a significantly suboptimal resolution, which is not the case for the maps considered).

Contour plots for $\alpha$ and $\l_0$ are shown in Fig.~2. They have been obtained considering a Gaussian likelihood function $ {\mathcal{L}}=\exp{(-\chi^2/2)}$ and 
employing Bayesian statistics. \footnote{We acknowledge the use of sampling routines and an analysis tools of the CosmoMC package~\cite{Lewis:2002ah}}
We assume flat priors for all parameters, so the posterior probability is basically given by the likelihood and the method corresponds to a maximum likelihood estimation where contours in Fig.~2 are given by $\Delta \chi^2=\chi^2-\chi^2_{bf}$, where $\chi^2_{bf}$ is the $\chi^2$ of best-fit.
This procedure does not imply parameters outside the confidence regions have necessarily a bad absolute $\chi^2$, but such technique is more reliable than standard frequentist schemes when deriving preferred values of parameters for a given model, which is the aim of this work (for a recent review on the topic, see, e.g.,~\cite{Hogg:2010yz}).

\begin{figure}[t]
 \begin{minipage}[htb]{0.5\textwidth}
   \centering
   \includegraphics[width=\textwidth]{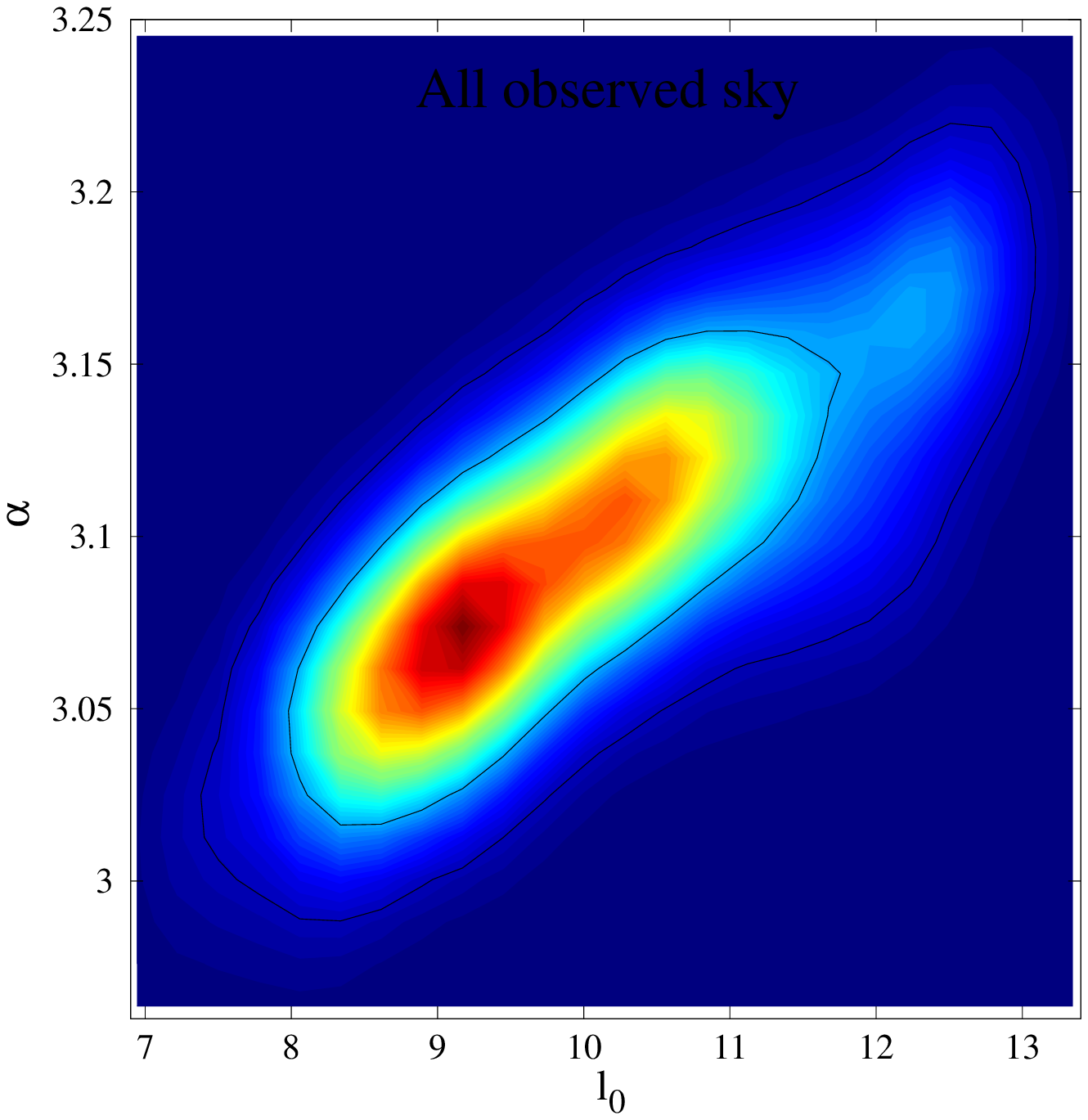}
 \end{minipage}
 \ \hspace{2mm} \
 \begin{minipage}[htb]{0.5\textwidth}
   \centering
   \includegraphics[width=\textwidth]{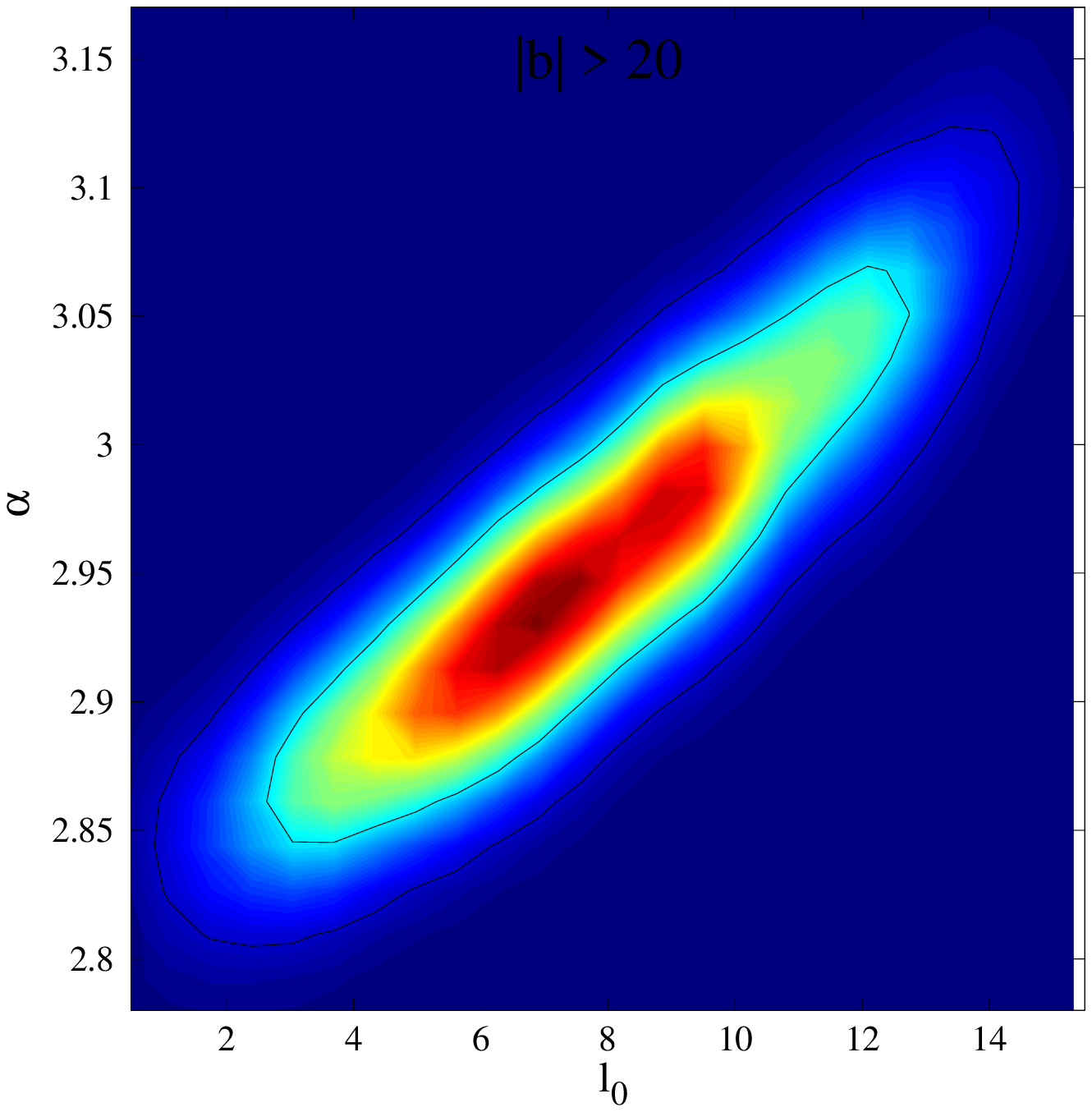}
 \end{minipage}
    \caption{Contour plot of $\alpha$ and $\l_0$ for the mean likelihood and the 68\% and 95\% C.L. regions (black lines). The fit has been performed on all available data in all maps (left) and applying a cut for $b<20$ (right). See text for details.
}
\label{fig:contour}
 \end{figure}

\section{Results and Discussion}
By fitting $\cl$'s from all observed sky, we derive $\alpha=3.1\pm 0.2$ and $\l_0=10\pm3$ at 95\% C.L.; considering high-latitudes data only, $\alpha=2.9^{+0.3}_{-0.1}$ and $\l_0=5^{+10}_{-4}$ are obtained. The best-fit $\chi^2$ is found to be $\chi^2_{bf}=63.6$ (with 80 bins considered) and $\chi^2_{bf}=65.3$ (with 64 bins), respectively (we note also that fitting with a single power-law, one finds $\chi^2_{bf}=87.2$ and $\chi^2_{bf}=66.2$; therefore such model, which however would lead to an even flatter spectrum, is disfavoured with respect to the one considered here).
The best fit of $\alpha$ is slightly shallower at high-latitudes but compatible at 1-$\sigma$ level.
One might expect $\l_0$ to be found slightly smaller at high latitudes which is indeed the case for best-fit values, although the two results are again compatible at 1-$\sigma$. 
We also check that best fits of flat components $\cl^{s}$ are consistent with subdominant contributions in the whole multipole range $\l=[5,100]$, which (together with the very mild variations of $\cl$'s from full-sky to Galactic plane cutoff) confirms assumptions about point sources discussed in previous Sections. 

We can confidently affirm that our results implies $\gamma=4-\alpha\gtrsim 0.7$.
As discussed above, the multipole of transition $\l_0$ is related to the geometry of the turbulence (namely, corresponds to the angle above which turbulences start to be less and less correlated since lines of sight pass through turbulences of separated turbulent regions) and its fit provides the size of the turbulent halo in terms of the turbulence scale. We have $d_{max}=\l_0\cdot L/\pi\lesssim 0.5 \frac{L}{100\,{\rm pc}}$~kpc (where 100 pc is a typical scale for $L$ as motivated below).

The spectral index and halo scale of turbulence are extremely important parameters in the description of propagation of high-energy cosmic-ray particles in the Galaxy. Indeed, for isotropic turbulences (i.e., same assumption of above analysis) and in the quasi-linear approximation, the diffusion tensor simplifies to the scalar expression~\cite{Berezinskii:1990}:
\be
D=\frac{v\,r_g}{12\,\pi}\frac{B^2}{k_{res}\,\tilde P(k_{res})}=\frac{v\,r_g^\gamma}{3\,(1-\gamma)}\frac{B^2}{k_L^{1-\gamma}\,\delta B_L^2}\,,
\label{eq:D1}
\ee
where $r_g=1/k_{res}=R/B$ is the gyroradius (with $R$=particle rigidity), $\tilde P(k)\propto k^{\gamma-2}$ is the turbulence power spectrum (again assumed to be a power-law), $k_L$ is a wavenumber where the random magnetic field assume the value $\delta B_L$, and the power spectrum is normalized through $\int^{\infty}_{k_L}\tilde P(k)=\delta B_L^2/4\pi$.

It is common practice to reduce Eq.~\ref{eq:D1} to $D=\beta D_0\,R^\gamma$, where $\beta=v/c$. The coefficient $D_0$ is often assumed to be constant up to a vertical scale $z_h$ after which free escape of particles is considered. Or, similarly, $z_h$ can indicate the scale at which the diffusion coefficient becomes large (e.g., $D_0\propto \exp{(z/z_h)}$~\cite{Regis:2009md}).

Recent derivations of spectral index $\gamma$ from cosmic-ray data can be split into two categories. If data are interpreted assuming a low convection wind (i.e., null or linearly increasing in the vertical direction), ratios of secondaries to primaries can be typically well-fitted by diffusion (plus reacceleration) models with Kolmogorov ($\gamma=1/3$)~\cite{Trotta:2010mx} or Kraichnan ($\gamma=1/2$)~\cite{DiBernardo:2009ku} spectrum of turbulence. On the other hand, when convective effects are introduced, flatter turbulence spectra ($\gamma=0.75-0.85$) are favoured~\cite{Maurin:2001sj,Putze:2010zn,Maurin:2010zp}. More generally, CR models foreseeing processes other than diffusion to be responsible for the flattening of $B/C$ at high-energy (as e.g., convection or sources spectrum) allow large value of $\gamma$.

The analysis in this paper supports this possibility.
Note that our result is in agreement with previous works~\cite{Tegmark:1995pn,Bouchet:1999gq,Baccigalupi:2000pm,Giardino,LaPorta:2008ag,Cho:2010kw}, all finding, with different uncertainties, a spectral index $\gamma\sim 1$ for $\l\lesssim$ 100.

This interpretation of our result is, however, based on a crucial and not obvious assumption.
The relevant turbulence scale for scatterings of CR particle with hydromagnetic waves is the gyroradius of the particle (e.g., for GeV CR particles, $r_g\sim 10^{12}$ cm). 
This is much smaller than the scales of turbulence probed by our analysis.
We have in mind a scenario where supernovae explosions and stellar winds inject turbulent energy at scales $\sim L$; then such energy is transferred through a direct cascade mechanism (with no dissipation) towards smaller scales. The resulting small-scale turbulence is expected to show the same spectral index of larger turbulence. 
However, one can argue that a straightforward extrapolation of the power spectrum down to the very small scales of CR scatterings can be a hazard.

Observations of turbulent magnetic field at very small scales (i.e., of order of $10^{12}$ cm) are lacking. On the contrary, fluctuations of thermal electron density in the nearby interstellar medium have been shown to be well approximated by a Kolmogorov spectrum~\cite{Armstrong:1995zc}, for scales ranging from $10^{10}$ cm to $10^{20}$ cm.
Then it is commonly postulated that magnetic fields are frozen into the ionized interstellar medium and so magnetic turbulence should show the same power-spectrum~\cite{Han:2009ts}; however this possibility, although reasonable, needs to be proven. E.g., as a counter-example, at scales $\gtrsim 10^{19}$ cm, observational evidences suggest a significantly flatter spectrum for the magnetic component of the turbulence, as we will comment below.

Turbulence observations most often rely on Faraday rotation measurements (RM). From RM only, on the other hand, it is not possible to separate fluctuations in electron density from magnetic field inhomogeneities (since RM $= \int B\cdot n_e dl$). 
By combining emission measure (EM $=\int n_e^2 dl$) with RM of polarized background radio sources, Ref.~\cite{MinterSpangler} estimated the magnetic turbulence at angular scales $1^\circ<\theta<10^\circ$ (corresponding to scales between $\sim$ 10 pc and 100 pc and roughly overlapping with the scales $\l<100$ considered in this work) finding a spectral shape significantly flatter than in a Kolmogorov case, and, more precisely, $\alpha=2.7\pm0.1$, which is compatible at the 2-$\sigma$ level with above results (and, in particular, with
$\alpha$ is fitted from high-latitudes data, which is the most directly comparable case).\footnote {The quantity derived in Ref.~\cite{MinterSpangler} was the structure function $D\propto \theta^\mu$. For small angles, the spectral index $\mu$ is 
simply related to the spectral index of $C_\ell$ by $\mu=\alpha-2$~\cite{Cho:2010kw}.} 
At smaller scales, EM data were not available; using RM data only the turbulence was found to follow a similar power spectrum up to scales $\sim 4$ pc, below which a confident distinction between different scenarios is quite hard (but with Kolmogorov spectrum being perfectly viable down to the limit of the survey $\sim$ 0.03 pc).
Although the analysis in~\cite{MinterSpangler} relies on some assumptions (e.g., same spectral index for electron and magnetic fluctuations) and is limited to a small high-latitude region in the outer Galaxy (while significant variations of RM structure function across the sky have been reported~\cite{Han:2004aa,Haverkorn:2008tb,Stil:2010np}), the rough agreement with this work is very encouraging.\footnote{Note that rotation measures are sensitive to the line-of-sight component of the magnetic field in thermal gas, while synchrotron data probe the perpendicular component of the field illuminated by cosmic rays.
However, under our assumption of isotropic and homogeneous turbulence, the two can be directly compared.}
On larger scales, between  0.5 and 15 kpc (i.e., above the strictly turbulent domain which is $\lesssim L$, so possibly probing magnetic fluctuations of different origin), Ref.~\cite{Han:2004aa} derived the magnetic spectrum from the ratio of RM and dispersion measure (DM $=\int n_e dl$) of pulsars in the Galactic disk, finding a nearly flat spectrum with $\alpha-2\sim 0.4$.

From a theoretical point of view, it is not completely clear if a Kolmogorov-like spectrum should be expected. Many numerical simulations of 3D magneto-hydrodynamic
turbulence have reported a shallower power spectrum, and the issue is currently under debate (for recent reviews see, e.g.,~\cite{Cho:2002qn,Brandenburg:2009tf} and references therein). 

To summarize this part, the state of the art about magnetic field turbulence is that from the observational data we have (including this work), which sample turbulence only at scales above few pc, the power spectrum is found to be much flatter than in the Kolmogorov case.
At smaller scales, there are hints for a Kolmogorov power spectrum, but data which could lead to conclusive proofs or to conclusive exclusions of other models are lacking.

So we can take two perspectives. We can assume results found in this work to hold all the way down to very small scales, and the above discussion on consequences for CR models follows. Conversely, we can assume a Kolmogorov power spectrum at small scales, and take the results as an evidence for a transition between two regimes, namely, a departure from self-similarity for the observed spectrum. 
For example, Ref.~\cite{MinterSpangler} considered the possibility that the Kolmogorov turbulence is in form of thin sheets which for large angular scales effectively appears as 2D turbulence. In this scenario, going from small to large scales, a transition from a 3D to a 2D turbulence regime would occur, where a 2D Kolmogorov power spectrum $\alpha=8/3$ is not far from our best-fit values. Note that this uncertainty (namely, a degeneracy between spectral and spatial properties of turbulence) is a general caveat in these analyses and is given by the fact that we are able to observe only 2D projections of underlying 3D spectra.

For what concerns the halo height $z_h$, it is not easily constrained by CR data since they typically depends on the ratio $D_0/z_h$, rather than on $z_h$ only. The strongest bounds come from `radioactive clocks', namely, unstable secondaries, as mentioned in the Introduction, with the most precise measurements being, at present, $^{10}$Be/$^9$Be.
Neglecting local effects, they suggest a diffusive zone significantly thicker than the Galactic disc, i.e., $z_h\gtrsim4$ kpc~\cite{Trotta:2010mx,Maurin:2001sj,Putze:2010zn}; this is also supported by observations of significantly extended magnetic structures in external galaxies (for a recent analysis of a spiral disk galaxy, see, e.g.,~\cite{Heesen:2008cs}.
  
Another important observable for deriving magnetic halo properties 
is the Galactic total synchrotron intensity (see, e.g.,~\cite{Sun:2009bg}).
However, in this case, the disentanglement of spatial 
properties of magnetic fields from cosmic-rays distribution properties, 
and of different magnetic components (regular and turbulent) 
themselves is very challenging.

Although $d_{max}$ cannot be immediately identified with $z_h$, it represents a rough estimate, in particular if extracted from observations at high-latitudes, see above.
Such estimate crucially depends on assumptions about outer scale of
turbulence. The dominant source of turbulence is believed to be supernova
remnants (and superbubbles), injecting energy in the surrounding interstellar medium 
on scales $L\sim100$ pc (for a review, see, e.g.,~\cite{MacLow:2003mz}).
Therefore, we infer a significantly thinner halo $z_h\lesssim1$ kpc, in apparent conflict with CR data.
 
The disagreement could be alleviated if most SNRs are clustered 
with supershells of much larger extent than 100 pc (basically, $z_h\sim4$ kpc would require $L\sim1$ kpc).
However, although many H{\small I} supershell have been observed in the Galaxy~\cite{McClureGriffiths:2002ic} (and in external galaxies as well) 
extending up to kpc-scales (and presumably crated by SN explosions),
it is unlikely that the mean characteristic outer scale of turbulence 
(i.e., the scale at which SNR energy is transferred to interstellar medium) 
is significantly larger than 100 pc, which follows from both simulation (e.g.,~\cite{deAvillez:2005yd}) and observational (e.g., discussion in \cite{Haverkorn:2008tb}) arguments. 

On the other hand, the local environment can significantly affect abundances of radioactive species, since they travel short distances before decaying (of order of 100 pc). In particular, an underdensity of a similar size surrounding the Sun can exponentially increase the amount of decays (lowering the local ratio)~\cite{Donato:2001eq}. Therefore the model of the local interstellar medium is crucial for the determination of $z_h$ from unstable secondaries.
Current observations~\cite{Redfield:2009} do not allow a very detailed description of the Local Bubble to precisely estimate this effect~\cite{Strong:2007nh,Putze:2010zn}.
However, assuming a simple and reasonable underdensity of $r_h\sim 100$ pc, bounds on $z_h$ can be strongly relaxed and $z_h\lesssim1$ kpc becomes fully viable~\cite{Donato:2001eq,Putze:2010zn}.
Moreover, Ref.~\cite{Maurin:2001sj} found that the diffusive zone in models with large $\gamma$ is constrained to be rather small.

A consistent (but not conventional) picture seems to emerge and deserves further investigation.

\section{Conclusion}
We have presented an analysis of the angular power spectrum in five radio maps at multipoles $5<\l<100$.
At these scales, the $\cl$'s are a tracer of fluctuations in the Galactic synchrotron emission, since contributions from sources or other diffuse components are subdominant.
At frequency below few GHz, large-scale anisotropies in synchrotron radiation are mainly induced by magnetic turbulence, 
while the non-thermal electron population involved in the emission has a rather smooth spatial distribution.
We assumed a simple isotropic and power-law spectrum of turbulence $P(k)\sim k^{-\alpha}$. 
The theoretical angular power spectrum thus scales as $\cl\propto \l^{-1}$ before $\l_0 \sim 30\,d_{max}/$kpc and $\cl\propto \l^{-\alpha}$ afterwards, where $d_{max}$ can be considered in first approximation as an estimate for the boundary of the turbulent diffusive halo. 
Results of fits are mildly dependent on latitudes. The spectral index is found to be $\alpha\sim 3$, much flatter than Kolmogorov ($\alpha=11/3$) or Kraichnan ($\alpha=7/2$) cases often considered in the literature. Moreover, data favour a very thin halo, $z_h\lesssim1$ kpc. 

If extrapolated to smaller scales, such Galactic turbulence scenario is consistent with cosmic-ray data, but pointing towards non-conventional models where convective winds play a significant role. 
On the contrary, assuming Kolmogorov turbulence to hold at scales comparable to the gyroradius of Galactic CR particles, our results can imply a non-trivial mechanism for transferring turbulent energy from large to small scales.

\section*{Acknowledgements}
We would like to thank F.~Donato and N.~Fornengo for very useful comments, and the anonymous referee for a detailed report which helped in substantially improving the paper.
We acknowledge facilities of the Centre for High Performance Computing, Cape Town.
This work is supported by Research Grants funded jointly by Ministero dell'Istruzione, dell'Universit\`a e della Ricerca (MIUR), by
Universit\`a di Torino and by Istituto Nazionale di Fisica Nucleare within the {\sl Astroparticle Physics Project} 
(MIUR contract number: PRIN 2008NR3EBK; INFN grant code: FA51).

\end{document}